\documentclass{article}
\usepackage[utf8]{inputenc}
\usepackage{amsmath}
\usepackage{amsfonts}
\usepackage{amssymb}
\usepackage{epstopdf}
\usepackage{graphicx}

\title{Nodal-line dynamics via exact polynomial solutions for coherent waves traversing aberrated imaging systems}
\author{David M. Paganin*, Mario A. Beltran** \& Timothy C. Petersen*}

\begin{document}

\maketitle
*School of Physics and Astronomy, Monash University, Australia; **School of Science, RMIT University, Australia and EMPA Swiss Federal Laboratories for Materials Science and Technology, Switzerland

\begin{abstract}
We obtain exact polynomial solutions for two-dimensional coherent complex scalar fields propagating through arbitrary aberrated shift-invariant linear imaging systems.  These are used to model nodal-line dynamics of coherent fields output by such systems. 
\end{abstract}

Quantized phase vortices occur in many physical systems described by complex scalar waves (Allen et al., 2001a).  Examples include the angular-momentum eigenstates of the hydrogen atom (Berry, 2001), phase vortices nucleated in the focal volume of a coherently illuminated lens (Boivin et al., 1967), electron vortices nucleated upon passage of a coherent electron plane wave through a crystal (Allen et al., 2001a), quantized phase vortices in Bose--Einstein condensates (Groszek et al., 2016) and the fractal tangle of vortices associated with coherent light speckle (O'Holleran et al., 2008).

The existence and the stability of phase vortices can be studied using a topological argument based on the continuity and single-valuedness of the complex wavefunction (Dirac, 1931).  A consequence is conservation of topological charge (phase winding number about a nodal line) for the nodal-line network in any three-dimensional subspace of $(x,y,\tau_1,\tau_2,\cdots)$ which coordinatizes a single-valued continuous complex wavefunction $\Psi$, where $(x,y)$ are transverse spatial coordinates and $(\tau_1,\tau_2,\cdots)$ are control parameters which can include, but are not limited to, propagation distance $z$ and time $t$ (Paganin, 2006).  Conservation of topological charge implies the possibility of critical-point explosions (Freund, 1999), where high-order topological charges and their associated nodal lines decay into a series of smaller-charge vortices. Nodal lines in each three-dimensional subspace of $(x,y,\tau_1,\tau_2,\cdots)$ form a connected network of one-dimensional lines that can form closed loops, knots, extend to infinity, end on surfaces where an underlying potential is discontinuous, or split/fuse at junctions/vertices that conserve topological charge.   

Such topological reactions of nodal-line networks, which thread the cores of phase-vortex structures, may be studied using exact finite-order polynomial solutions to the differential equations governing $\Psi(x,y,\tau_1,\tau_2,\cdots)$ (Nye \& Berry, 1974; Dennis et al., 2011).  Exact polynomial wavefunctions are an often convenient {\emph {local} description of an optical field (Nye \& Berry, 1974; Dennis et al., 2011), notwithstanding the divergent behavior of such solutions far from a compact region of interest.  

We consider exact finite-order forward-propagating polynomial fields passing through arbitrary linear imaging systems.  We construct a general polynomial wave valid for arbitrary aberrations, and apply it to four special cases of nodal-line dynamics.

With reference to Fig.~\ref{fig:Schematic}, consider a coherent scalar optical field whose spatial part may locally be described by a complex wavefunction $\Psi(x,y)\equiv\Psi({\bf r})$, over a two-dimensional $x-y$ plane.  This field is input into a two-dimensional shift-invariant coherent linear imaging system.  The resulting output is (Allen et al., 2001b; Paganin \& Gureyev, 2008):
%
\begin{eqnarray} 
\Psi \left ( \textbf{r}\mid \left \{ C_{mn} \right \} \right ) =\frac{1}{2\pi}\iint_{-\infty}^{\infty}d\textbf{k}_{\textbf{r}}\widehat{\Psi}( \textbf{k}_{\textbf{r}}\mid \left \{ C_{mn}\right \}=0) 
\nonumber \\
 \times \exp \left [ i \sum_{m=0}^{\infty} \sum_{n=0}^{\infty}C_{mn}k^{m}_{x}k^{n}_{y}+i\textbf{k}_{\textbf{r}}\cdot \textbf{r}\right ]. 
\label{ABBERATED_LSI_SYSTEM}
\end{eqnarray}
%
\begin{figure}[htbp]
\centering
\fbox{\includegraphics[width=\linewidth]{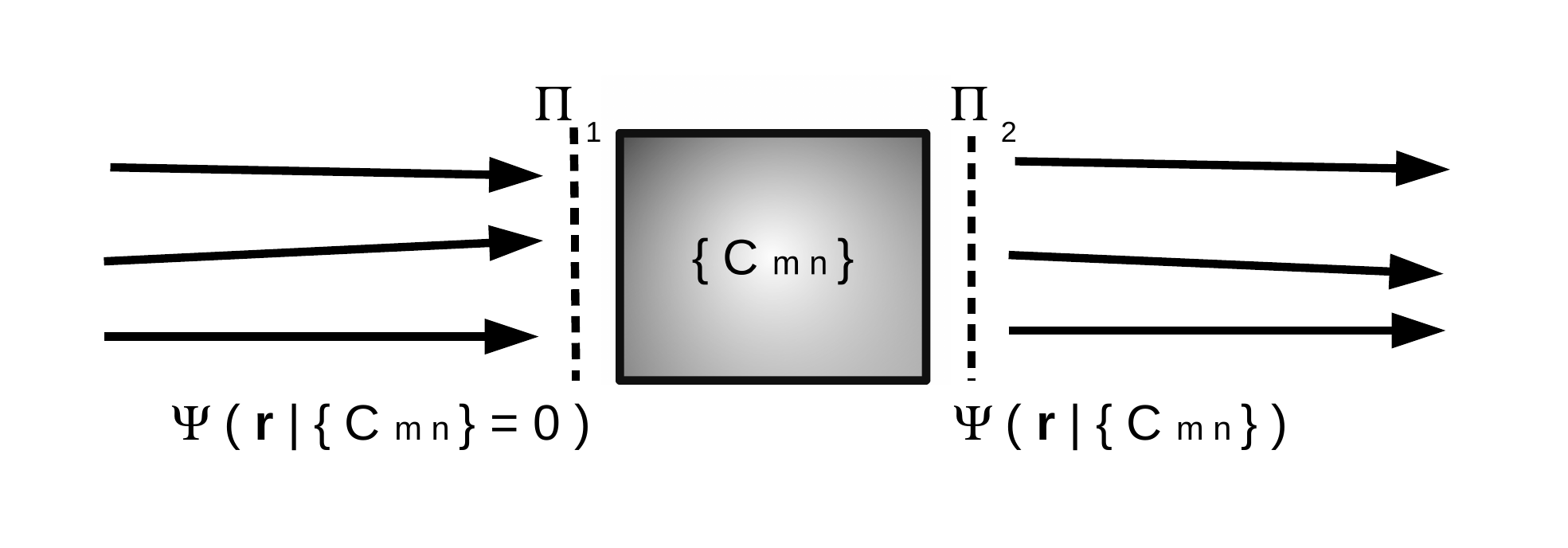}}
\caption{Forward-propagating field $\Psi({\bf r} | \{C_{mn}\}=0)$ enters a linear shift-invariant imaging system over plane $\Pi_1$.  The field is transmitted by the system characterized by aberration coefficients $ \{ C_{mn} \} $, to give $\Psi({\bf r} | \{C_{mn}\})$ over output plane $\Pi_2$.} 
\label{fig:Schematic}
\end{figure}

\noindent Here, the set of complex control parameters $\{ C_{mn} \}$ (``aberration coefficients'') specifies the state of the imaging system, $\widehat{\Psi}( \textbf{k}_{\textbf{r}}\mid \left \{ C_{mn}\right \}=0)$ is the Fourier transform of $\Psi({\bf r} | \{C_{mn}\}=0)$ with respect to ${\bf r}\equiv (x,y)$, and ${\bf k}_{\bf r}\equiv (k_x,k_y)$ are the corresponding Fourier coordinates, under the Fourier-transform convention:
%
\begin{subequations}
\begin{align}
\widehat{\Psi}( \textbf{k}_{\textbf{r}}) =\frac{1}{2\pi}\iint_{-\infty}^{\infty}d\textbf{r}~\exp(-i\textbf{k}_{\textbf{r}}\cdot \textbf{r})\Psi(\textbf{r}) \label{FOURIER_CONVENTION_A}, \\
 \Psi(\textbf{r}) =\frac{1}{2\pi}\iint_{-\infty}^{\infty}d\textbf{k}_{\textbf{r}}~\exp(i\textbf{k}_{\textbf{r}}\cdot \textbf{r})\widehat{\Psi}( \textbf{k}_{\textbf{r}}).
 \label{FOURIER_CONVENTION_B}
\end{align}
\end{subequations}
Each coefficient can be written as $C_{mn}=C_{mn}^{(R)}+ i C_{mn}^{(I)}$, where $C_{mn}^{(R)}$ and $C_{mn}^{(I)}$ are both real.  We speak of $C_{mn}^{(R)}$ and $C_{mn}^{(I)}$ as {\em coherent aberrations} and {\em incoherent aberrations}, respectively.  We impose the limitation that $C_{mn}^{(I)}\ge 0~{\textnormal{for all}}~(m,n)$ and $C_{mn}^{(I)}= 0$ if either or both of $(m,n)$ are odd, ensuring that incoherent aberrations may exponentially dampen $\widehat{\Psi}( \textbf{k}_{\textbf{r}}\mid \left \{ C_{mn}\right \}=0)$ at any particular spatial frequency $\textbf{k}_{\textbf{r}}$, but can never exponentially amplify it.  This amounts to assuming no gain media to be present.   

Apply the Fourier derivative theorem in reverse to Eq.~(\ref{ABBERATED_LSI_SYSTEM}), to give the following solution to the boundary-value problem of determining the output field corresponding to a specified ${\bf C}\equiv \left \{ C_{mn}\right \}$,  given the aberration-free input field:
%
\begin{eqnarray} 
\Psi \left ( \textbf{r} \mid {\bf C} \right ) = \exp \left[i \sum_{m=0}^{\infty} \sum_{n=0}^{\infty} \frac{C_{mn}}{i^{m+n}}  \frac{\partial^m}{\partial x^m} \frac{\partial^n}{\partial y^n}\right] \Psi \left ( \textbf{r} \mid {\bf C} = {\bf 0}\right ). 
\label{FORMAL_SOLUTION_WITH_ALL_ABBS}
\end{eqnarray}
%
Assume the input field $\Psi \left ( \textbf{r} \mid {\bf C} = {\bf 0}\right )$ is sufficiently well behaved, within a compact region of interest, to be locally well described by a finite-order Maclaurin expansion:
%
\begin{eqnarray} 
\Psi(\textbf{r} \mid {\bf C} = {\bf 0})=\sum_{p = 0}^{M} \sum_{q=0}^{N} \Theta_{pq} x^p y^q. \label{INPUT_POLYNOMIAL_FIELD}
\end{eqnarray}
%
\noindent Here, $M$ and $N$ are finite positive integers; each $\Theta_{pq}$ is complex.

Set $C_{00}=0$, which implies a trivial loss of generality since (i) $C_{00}^{(R)}$ corresponds to a global phase factor that has no effect on measured intensities; (ii) $C_{00}^{(I)}\ge 0$ gives a global attenuation which has no effect on relative intensities.  We also assume $C_{10}^{(R)}=C_{01}^{(R)}=0$, since the Fourier shift theorem implies that this merely generates transverse shifts in the output field.  

If Eq.~(\ref{INPUT_POLYNOMIAL_FIELD}) is substituted into Eq.~(\ref{FORMAL_SOLUTION_WITH_ALL_ABBS}), one obtains: 
\begin{eqnarray} 
\Psi(\textbf{r} \mid {\bf C}) = \quad\quad\quad\quad\quad\quad\quad\quad\quad\quad\quad\quad\quad\quad\quad\quad\quad\quad   \label{EXACT_POLYNOMIAL_SOLUTION_1} \\ \exp \left[i \sum_{m=0}^{M} \sum_{n=0}^{N} \frac{C_{mn}}{i^{m+n}}  \frac{\partial^m}{\partial x^m} \frac{\partial^n}{\partial y^n}\right]\sum_{p = 0}^{M} \sum_{q=0}^{N} \Theta_{pq} x^p y^q. \nonumber
\end{eqnarray}

\noindent This is a {\em finite-order exact polynomial solution} to the boundary-value problem of propagating Eq.~(\ref{INPUT_POLYNOMIAL_FIELD}) through a non-amplifying but otherwise arbitrary shift-invariant linear imaging system.  The finite order of the polynomial follows from the facts that: (i) power series expansion of the exponential will generate a linear combination of differential operators $(\partial^a/\partial x^a)(\partial^b/\partial y^b)$, with $a$ and $b$ being non-negative integers; (ii) differentiation can never increase the order of a finite polynomial. 

Since it acts on a finite-order polynomial, the exponential in Eq.~(\ref{EXACT_POLYNOMIAL_SOLUTION_1}) can be expanded as a Maclaurin series in which derivatives of order no higher than $(\partial^M/\partial x^M)(\partial^N/\partial y^N)$ appear.  One then obtains, utilizing combinatoric arguments similar to those given by Beltran et al.~(2015) in a different context,
%
%
\begin{eqnarray} 
\Psi(\textbf{r} \mid {\bf C}) =\left ( 1+ \sum_{m'=0}^{M} \sum_{n'=0}^{N} \frac{\Xi_{m'n'}}{i^{m'+n'}} \frac{\partial^{m'}}{\partial x^{m'} } \frac{\partial^{n'}}{\partial y^ {n'}} \right ) \sum_{p = 0}^{M} \sum_{q=0}^{N} \Theta_{pq} x^{p} y^{q}, \nonumber \\
\Xi_{m'n'}=\sum_{L}^{m'+n'}\sum_{J_{00}+J_{01}+...+J_{mn}=L}  \frac{i^L\mathfrak{I}~C^{J_{00}}_{00}C^{J_{01}}_{01}...C^{J_{mn}}_{mn}}{J_{00}!J_{01}!...J_{mn}!}, \quad\quad\quad
\label{COEFFICIENTS}
\end{eqnarray}
%
\noindent where $\mathfrak{I}$ is an indicator function equal to 1 if both $\sum_{m,n}mJ_{mn}=m'$ and  $\sum_{m,n}nJ_{mn}=n'$; it is 0 otherwise.  The indices $L=0,1,...,m'+n'$, $m=0,1,...,m'$, $n=0,1,...,n'$, $J_{mn}=0,1,...,m'+n'$ are non-negative integers. We have not assumed that $C_{00}=0$ when arriving at Eq.~(\ref{COEFFICIENTS}). This implies that for any $\Xi_{m'n'}$ there will be an infinite number of terms composed of $\left \{ C_{mn}\right \}$ combinations. Conversely, if one assumes $C_{00}=0$, $C_{01}=0$ and $C_{10}=0$ for the reasons stated previously, each $\Xi_{m'n'}$ reduces to finite series. 



Continuing with Eq.~(\ref{COEFFICIENTS}), interchange inner and outer double sums, truncate the resulting inner double-sum upper limits to exclude vanishing terms in the summand, then perform the differentiations on the summand explicitly.   Thus:
%
%
\begin{eqnarray} 
\nonumber\Psi(\textbf{r} \mid {\bf C}) =  \sum_{p,q,m',n' = 0} \mathcal{M}_{m',n'}^{p,q} x^{p} y^{q},  \quad  \nonumber \\ 
\mathcal{M}_{m',n'}^{p,q} = \left (  1+\frac{p!q!}{(p-m')!(q-n')!x^{m'}y^{n'} }\right )\frac{\Xi_{m'n'} \Theta_{pq}  }{i^{m'+n'}}.   \label{EXACT_POLYNOMIAL_SOLUTION_3}  
\end{eqnarray}

\noindent Here, $\sum_{p,q,m',n' = 0}$ implies a quadruple sum.  The above expression could be compacted into a double summation with the same form as Eq.~(\ref{INPUT_POLYNOMIAL_FIELD}).  One could also calculate exact polynomial expressions for the transverse current density, angular momentum density {\em etc.}  These expressions will not be given here.

Instead, turn attention to {\rm vortical} forms of the input polynomial wavefield in Eq.~(\ref{INPUT_POLYNOMIAL_FIELD}).  Suppose the input to contain an integer number $N_+$ of embedded vortices at specified locations $(x_j^+,y_j^+)$ with topological charges given by the positive integers $P_j$, $j=1,2,\cdots,N_+$, together with an integer number $N_-$ of embedded anti-vortices at specified locations $(x_l^-,y_l^-)$ with topological charges given by negative one times the positive integers $Q_l$, $l=1,2,\cdots,N_-$.  A finite-order input polynomial wavefield with this specified vortex--anti-vortex structure is (Smith \& Gbur, 2016): 
%
\begin{eqnarray} 
\Psi(\textbf{r} \mid {\bf C} = {\bf 0})=\prod_{j=1}^{N_+}[x-x_j^+ + i(y-y_j^+)]^{P_j} \nonumber \\  \times\prod_{l=1}^{N_-}[x-x_l^- - i(y-y_l^-)]^{Q_l	}. \label{INPUT_POLYNOMIAL_FIELD_VORTICAL}
\end{eqnarray}

Related polynomial wavefields can be generated through multiplicative or additive perturbation via linear combinations of real polynomials in $x-y$. Such perturbations can change the topology or break the symmetry of nodal lines in the propagated field.  For a chosen input vortex--anti-vortex configuration, one can use the above expression with chosen perturbing polynomials to compute the corresponding coefficients $\Theta_{pq}$ in Eq.~(\ref{INPUT_POLYNOMIAL_FIELD}).  Substitution into Eq.~(\ref{EXACT_POLYNOMIAL_SOLUTION_3}) then yields exact polynomial solutions for propagating the specified input vortical field through any non-amplifying two-dimensional shift-invariant linear imaging system.  Such exact expressions can be used to study nodal line dynamics, a topic to which we now turn.  

It is natural to consider a continuous family of states ${\bf C}(\tau)$ of the imaging system (Allen et al., 2001b), where $\tau\ge 0$ is a real parameter, with ${\bf C}(\tau=0)={\bf 0}$.  One can then consider each $C_{mn}$ to be a continuous function of $\tau\ge 0$ in any of the preceding expressions, which remain otherwise unchanged.  For simplicity, however, in the four special cases of nodal-line dynamics which follow, we restrict ourselves to varying a single particular coherent or incoherent aberration coefficient.

{\em 1: High-order-vortex decay via spherical aberration}

Our first special case is a perturbed doubly charged vortex at the origin, corresponding to the choice $(x+iy)^2 [1+A(x^2-y^2)]$ for the input polynomial field.  Here, $A \ge 0$ is a real parameter which perturbs our special case of Eq.~(\ref{INPUT_POLYNOMIAL_FIELD_VORTICAL}) so as to break the rotational symmetry of the input field.  The input perturbed charge-two vortex is acted upon by an aberrated imaging system containing only coherent fourth-order spherical aberration:
\begin{equation}\label{CsEq}
C_{40}^{(R)}=C_{04}^{(R)}=C_{22}^{(R)}/2\equiv C_S,
\end{equation}
\noindent with all other aberration coefficients vanishing. The polynomial expression for the resulting output field $\Psi(x,y,C_S)$, as a function of spherical aberration $C_S$, is the following special case of Eq.~(\ref{EXACT_POLYNOMIAL_SOLUTION_3}):
%
\begin{eqnarray} 
\Psi(x,y,C_S)=(x+iy)^2[1+A(x^2-y^2)]+32iAC_S.
\end{eqnarray} 

Equating both the real and imaginary parts of the above expression to zero, one finds that, while there is a single charge-two vortex at the origin when $C_S=0$, for other values of $C_S$ an aberration-induced ``critical point explosion'' (Freund, 1999) has occurred and one instead has two charge-one vortices.  In the positive-$C_S$ half of $x-y-C_S$ space, these two vortices are located at $(x,y)=(\pm\sqrt{16 C_S A},\mp\sqrt{16 C_S A})$, becoming progressively more widely spaced as $C_S \ge 0$ increases.  Conversely, in the negative-$C_S$ half of $x-y-C_S$ space, one again has two vortices; they are located at $(x,y)=(\pm\sqrt{-16 C_S A},\pm\sqrt{-16 C_S A})$, again becoming progressively more widely spaced as $C_S \le 0$ becomes increasingly negative.  These nodal-line dynamics, which exemplify the instability with respect to perturbation of high-order vortices, are sketched in Fig.~\ref{fig:VortexDecay}.  

\begin{figure}[htbp]
\centering
\fbox{\includegraphics[width=\linewidth]{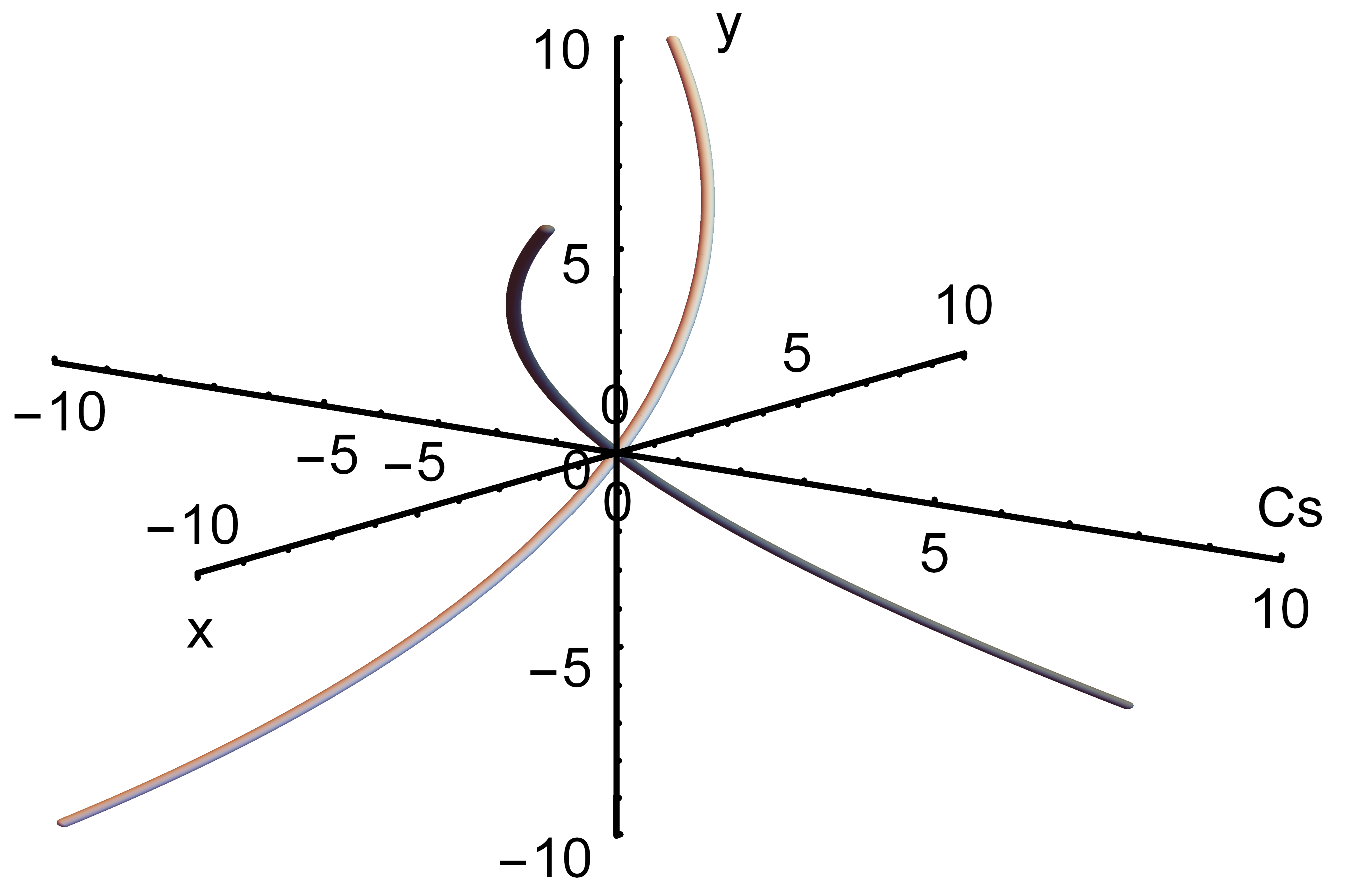}}

\caption{A charge-2 vortex, at the origin of the $x-y$ plane, decays to two charge-1 vortices upon passage through a linear imaging system with pure spherical aberration $C_S$.}
\label{fig:VortexDecay}
\end{figure}

{\em 2: Vortex--anti-vortex annihilation via incoherent blur}

Our second special case is a vortex--antivortex dipole, corresponding to $[x-(x_0/2)+iy][x+(x_0/2)-iy]$ for the input polynomial field.  Here, $x_0 \ge 0$ is the vortex--antivortex separation.  The input vortex--antivortex dipole is acted upon by an aberrated imaging system containing only Gaussian blur:
\begin{equation}
C_{20}^{(I)}=C_{02}^{(I)}=\sigma^2 \ge 0,
\end{equation}
\noindent with all other aberrations vanishing. The polynomial expression for the resulting output field $\Psi(x,y,\sigma)$, as a function of the Gaussian blur $\sigma$, is the following special case of Eq.~(\ref{EXACT_POLYNOMIAL_SOLUTION_3}):
%
\begin{eqnarray} 
\Psi(x,y,\sigma)=[x-(x_0/2)+iy][x+(x_0/2)-iy]-4\sigma^2.\quad\quad
\end{eqnarray} 

The vortex--antivortex pair in the dipole, separated by $x_0$ when $\sigma=0$, approach one another along an elliptic trajectory as $\sigma$ is increased above zero, with progressively-decreasing separation.  The pair, located at $
(x,y)=\left(\pm\sqrt{x_0^2/4-4\sigma^2},0\right)$, eventually coalesce when $\sigma=x_0/4$ -- see Fig.~\ref{fig:VortexDipoleDecay}.  The action of the imaging system, which in this case is diffusive, independently coarse-grains the real and imaginary parts of the complex wavefunction, thereby ``healing'' the screw-type Riemann-sheet tears in the phase via mutual vortex--antivortex annihilation.

\begin{figure}[htbp]
\centering
\fbox{\includegraphics[width=\linewidth]{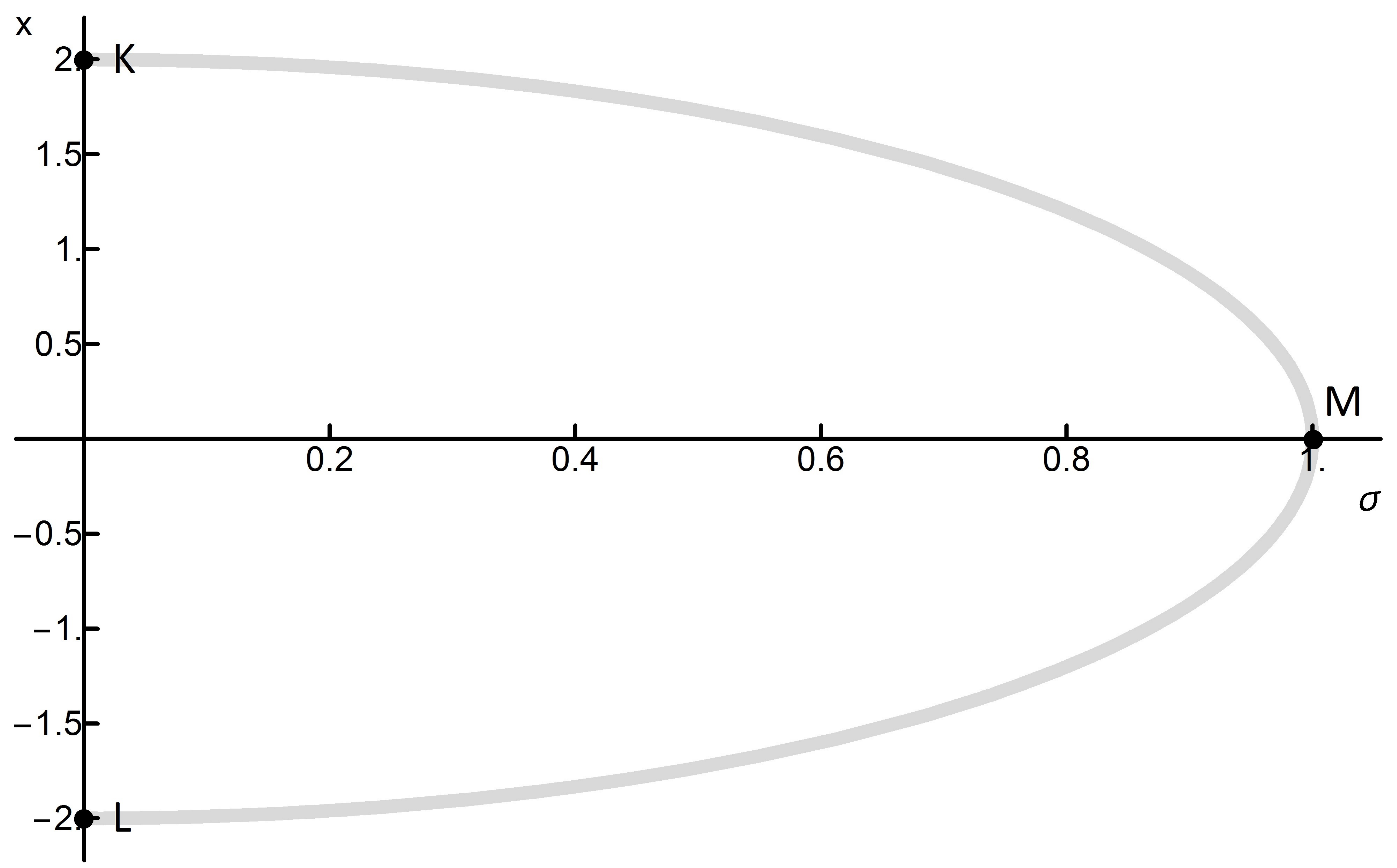}}
\caption{A vortex--antivortex dipole $KL$, with separation $x_0 = 4$ in the $x-y$ plane ($y$ axis not shown), is annihilated at $M$ by pure Gaussian blur $\sigma$.}
\label{fig:VortexDipoleDecay}
\end{figure}

{\em 3: Astigmatism-induced vortex quadrupole}

Consider the ring of intensity zeros, surrounding the Airy disc in the focal plane of a collapsing paraxial complex scalar wave that is truncated by a sharp circular aperture upstream of the focal plane.  This circle of focal-plane zeros around the central Airy disc is a one-dimensional nodal line with unit topological charge in the three-dimensional space occupied by the collapsing wave.  The ``domain wall'' phase shift of $\pi$ radians, as one crosses this topologically unstable nodal line along a path entirely contained within the focal plane, corresponds to half of the phase winding obtained when one traverses a full closed circuit that encloses the said nodal line.  When the illuminated lens is subsequently aberrated in a manner that does not possess rotational symmetry, the focal-plane nodal-line ring may become deformed or puckered, piercing the focal plane in a stitch-like manner, creating a ring of alternating-charge vortices (Walford et al., 2002). 

Inspired by the topology of this process, consider the following ``input'' wavefunction comprising a unit-radius phase-domain-wall circle of zeros, whose $B=0$ rotationally symmetric case is broken when the real parameter $B$ exceeds zero: 
%
\begin{eqnarray} 
\Psi(x,y,B)=(x^2+y^2-1)(1+Bx^2),  \quad B \ge 0.
\label{eq:InputNodalLineRing}
\end{eqnarray} 
%
\noindent Let $C_{11}^{\textrm{(R)}}\equiv A$ be an astigmatism aberration, with all other aberration coefficients assumed to vanish.  Equation~(\ref{EXACT_POLYNOMIAL_SOLUTION_3}) then gives:
%
\begin{eqnarray} 
\Psi(x,y,B,A)=(x^2+y^2-1)(1+Bx^2)-2A^2B-4iABxy.
\end{eqnarray} 

If either or both of the symmetry-breaking term $B$ or the astigmatism $A$ vanish, one retains the unit-radius phase-domain-wall ring of zeros.  If both $A$ and $B$ are non-zero, the ring of zeros is deformed into a topologically stable vortex quadrupole.  This consists of two charge $+1$ vortices and two charge $-1$ vortices, both of which lie outside the unit circle, with one pair of like-signed vortices arranged along the $x$ axis at  
%
\begin{eqnarray} 
(x,y)=\left(\pm\sqrt{ 1 + \frac{\sqrt{(B+1)^2+8A^2B^2}-1  }{B}},0\right), \quad B>0, 
\end{eqnarray} 

\noindent and the other pair of opposite-signed vortices  arranged along the $y$ axis at $(x,y)=\left(0,\pm\sqrt{1+2A^2B}\right), B>0$ -- see Fig.~\ref{fig:PuckedNodalRing}. 

\begin{figure}[htbp]
\centering
\fbox{\includegraphics[width=\linewidth]{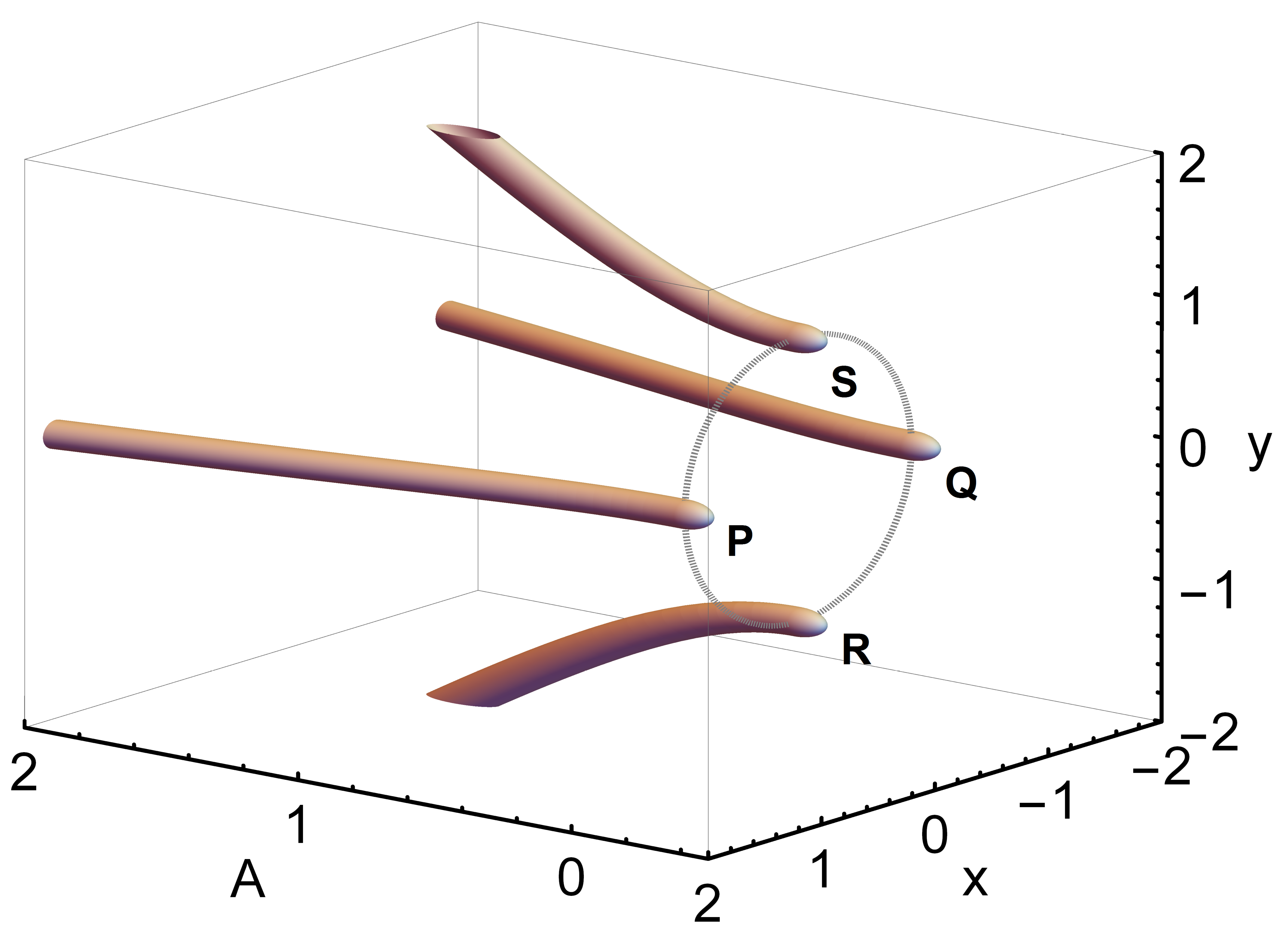}}
\caption{Topologically unstable domain-wall ring of zeros of input field in Eq.~(\ref{eq:InputNodalLineRing}), shown as gray loop, transforms to topologically stable vortex quadrupole $PQRS$ via astigmatism $A$.}
\label{fig:PuckedNodalRing}
\end{figure}

\newpage

{\em 4: An isolated trefoil knot in spherical aberration space}

Knots can be embedded in polynomial wavefields.  For example, Dennis et al.~(2010) theoretically devised and experimentally demonstrated a variety of  optical knots. Dennis et al.~(2010, 2011) define knots by substituting `Milnor maps' $\{u(r), v(r)\}$ into polynomial expressions describing braids.  Milnor maps define complex coordinates for the 3-sphere, for a particular stereographic projection, and are composed to wrap the braids onto a torus and thereby create knots.  For helical knots, such as the trefoil, this construction can be used to define polynomial wavefunctions in $x-y-z$ space with knotted nodal lines, if one excludes the denominators in the Milnor maps (which does not affect phase-field topology).  For example, the trefoil wavefunction is given by composing $q_{\textrm{trefoil}}(u,v) = u^2-v^3$ with the numerators $u(x,y,z)=(x^2+y^2-1)+2iz$ and $v(x,y,z)=2(x+iy)$.  

Remarkably, Dennis et al.~showed that a $z = 0$ section of this wave function recreates the trefoil vortical knot upon paraxial propagation over the distance $z$.  With $z$ representing the defocus aberration, we can show this by acting the coherent aberration, $C_{20}^{(R)}=C_{02}^{(R)}=z$, upon the $z = 0$ section representing the input wave function, $\Psi(x,y)=(x^2 + y^2-1)^2-8(x + i y)^3$, 
giving
\begin{eqnarray} 
\Psi(x,y,z)=( x^2 + y^2-1)^2 -8 (x + i y)^3 \nonumber \\ + 
 8i (2x^2 + 2y^2 -1) z - 32 z^2\end{eqnarray} 
in $x-y-z$ space.  This wavefunction is not the same as the wavefunction $q_{\textrm{trefoil}}(u(x,y,z), v(x,y,z))$, differing by an imaginary quadratic in $x-y$ plus a real $z^2$ term, but the vortices of both functions trace out the trefoil knot. 


To demonstrate a knot in {\em aberration} space, beyond defocus, we can attempt to recreate monomials missing from the output wavefunction when compared against $q_{\textrm{trefoil}}(u(x,y,z),v(x,y,z))$.  A suitable trial for an input trefoil wavefunction in $x-y-C_S$ space is $\Psi(x,y)=(x^2 + y^2 - \alpha)^4\beta - 8 (x + i  y)^3$. For spherical aberration defined by Eq.~(\ref{CsEq}), this input choice creates the $C_S$-propagated output wavefunction: 
\begin{eqnarray} 
\Psi(x,y,C_S)=-8 (x + i y)^3 - 
 2^{13} 3^2 C_S^2 \beta + (x^2 + y^2 - \alpha)^4 \beta \nonumber \\ +  
 2^7 3 i C_S (6 (x^2 + y^2)^2
    -6 (x^2 + y^2) \alpha + \alpha^2) \beta. \end{eqnarray}
Unlike all previous examples in this Letter, the roots of this wavefunction cannot be solved analytically, given the eighth order polynomial form.  However, roots can be found numerically.  Particular choices of $\alpha$ and $\beta$ were found to result in a trefoil knot, while other choices produced loops.  For example, setting $\alpha = 8$ and $\beta = 2/5$ gives the trefoil knot in Fig.~\ref{fig:CsKnot}.

\begin{figure}[htbp]
\centering
\fbox{\includegraphics[width=\linewidth]{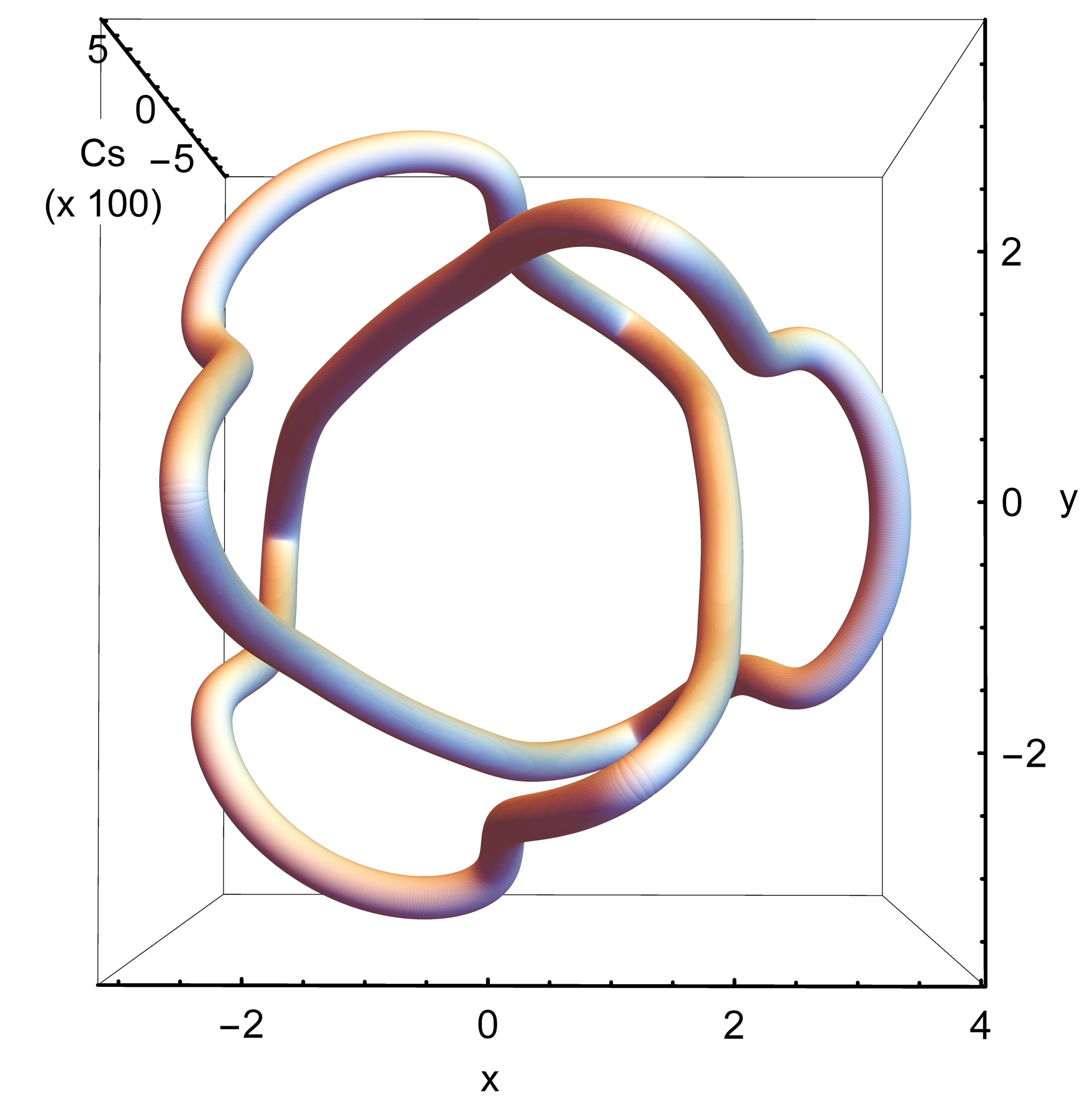}}
\caption{An isolated trefoil knot in $(x-y-C_S)$ space}
\label{fig:CsKnot}
\end{figure}


\section*{References}

L.J.~Allen, H.M.L.~Faulkner, M.P.~Oxley and D.~Paganin, \textnormal{``Phase retrieval and aberration correction in the presence of vortices in high-resolution transmission electron microscopy,''} Ultramicroscopy \textbf{88}, 85--97 (2001a).

L.J.~Allen, M.P.~Oxley and D.~Paganin, \textnormal{``Computational aberration correction for an arbitrary linear imaging system,''} Phys. Rev. Lett. \textbf{87}, 123902 (2001b).

M.A.~Beltran, M.J.~Kitchen, T.C.~Petersen and D.M.~Paganin, \textnormal{``Aberrations in shift-invariant linear optical imaging systems using partially coherent fields,''} Opt. Commun. \textbf{355}, 398--405 (2015).

M.V.~Berry, \textnormal{``Knotted zeros in the quantum states of hydrogen,''} Found. Phys. \textbf{31}, 659--667 (2001).

A.~Boivin, J. Dow and E.~Wolf, \textnormal{``Energy flow in the neighborhood of the focus of a coherent beam,''} J. Opt. Soc. Am \textbf{57}, 1171--1175 (1967).

M.R.~Dennis, R.P.~King, B.~Jack, K.~O'Holleran and M.J.~Padgett, \textnormal{``Isolated optical vortex knots,''} Nature Phys. \textbf{6}, 118--121 (2010).

M.R.~Dennis, J.R.~G\"{o}tte, R.P.~King, M.A.~Morgan and M.A.~Alonso, \textnormal{``Paraxial and nonparaxial polynomial beams and the analytic approach to propagation''}, Opt. Lett. \textbf{36}, 4452--4454 (2011).

P.A.M.~Dirac, \textnormal{``Quantised singularities in the electromagnetic field,''} Proc. Roy. Soc. Lond. A \textbf{133}, 60--72 (1931).

I.~Freund, \textnormal{``Critical point explosions in two-dimensional wave fields,''} Opt. Commun. \textbf{159}, 99--117 (1999).

A.J.~Groszek, T.P.~Simula, D.M.~Paganin and K.~Helmerson, \textnormal{``Onsager vortex formation in Bose-Einstein condensates in two-dimensional power-law traps,''} Phys. Rev. A \textbf{93}, 043614 (2016).

K.~O'Holleran, M.R.~Dennis, F.~Flossmann and M.J.~Padgett, \textnormal{``Fractality of light’s darkness,''} Phys. Rev. Lett. \textbf{100}, 053902 (2008).

J.F.~Nye and M.V.~Berry, {``Dislocations in wave trains,''} Proc. Roy. Soc. Lond. A \textbf{336}, 165--190 (1974).

D.M.~Paganin, {\em{Coherent X-Ray Optics}}, \textnormal{(Oxford University Press, Oxford, 2006)}.

D.M.~Paganin and T.E.~Gureyev, \textnormal{``Phase contrast, phase retrieval and aberration balancing
in shift-invariant linear imaging systems,''} Opt. Commun. \textbf{281}, 965--981 (2008).

M.K.~Smith and G.J.~Gbur, \textnormal{``Construction of arbitrary vortex and
superoscillatory fields,''} Opt. Lett. \textbf{41}, 4979--4982 (2016).

J.N.~Walford, K.A.~Nugent, A.~Roberts and R.E.~Scholten, \textnormal{``High-resolution phase imaging of phase singularities in the focal region of a lens,''} Opt. Lett. \textbf{27}, 345--347 (2002).

\end{document}